\documentclass[english,journal=jctce,manuscript=article]{achemso}
\usepackage[T1]{fontenc}
\usepackage[latin9]{inputenc}
\usepackage{array}
\usepackage{pmboxdraw}
\usepackage{multirow}
\usepackage{amsmath}
\usepackage{graphicx}

\makeatletter

\title{Minimally-corrected partial atomic charges for non-covalent electrostatic
interactions }
\author{Rebecca Efrat Hadad }
\affiliation{Fritz Haber Center for Molecular Dynamics, Institute of Chemistry,
The Hebrew University of Jerusalem, Jerusalem 91904 Israel}
\author{Roi Baer}
\affiliation{Fritz Haber Center for Molecular Dynamics, Institute of Chemistry,
The Hebrew University of Jerusalem, Jerusalem 91904 Israel}
\email{roi.baer@huji.ac.il}
\providecommand{\tabularnewline}{\\}

\SectionNumbersOn

\makeatother

\usepackage{babel}
\begin{document}
\begin{abstract}
We develop a new scheme for determining molecular partial atomic charges
(PACs) with external electrostatic potential (ESP) closely mimicking
that of the molecule. The PACs are the ``minimal corrections'' to
a reference-set of PACs necessary for reproducing exactly the tensor
components of the Cartesian zero- first- and second- molecular electrostatic
multipoles. We evaluate the quality of ESP reproduction when ``minimally
correcting'' (MC) Mulliken, Hirshfeld or iterated-Hirshfeld reference
PACs. In all these cases the MC-PACs significantly improve the ESP
while preserving the reference PACs' invariance under the molecular
symmetry operations. When iterative-Hirshfeld PACs are used as reference
the MC-PACs yield ESPs of comparable quality to those of the ChElPG
charge fitting method.
\end{abstract}

\section{Introduction}

Partial atomic charges (PACs), i.e. point charges placed on the nuclei
position of a molecule are often used in large-scale molecular mechanics
calculations to replace the detailed quantum mechanical charge distributions.
\cite{Lifson1968,Warshel1976,Allinger1989a,Field1990,Duffy2000,Politzer2013,Mei2015}
The model is extremely useful since by using them the long-range electrostatic
forces acting \emph{between }molecules can be expressed as a sum of
pairwise interactions, enabling a fast computation, important especially
as molecules jiggle around and rotate quite a lot during the course
of the simulation. The question of just how to determine PACs for
this purpose is critical. We argue that the most important constraint
is the exact reproduction of the low-order electrostatic moments (ESM),
the monopole $Q=e\int\rho\left(\boldsymbol{r}\right)d\boldsymbol{r}$,
which is the total charge of the system, the dipole $\mu_{i}=e\int\rho\left(\boldsymbol{r}\right)r_{i}d\boldsymbol{r}$
($i=x,y,z$) and the quadrupole moment $\Theta_{ij}=e\int\rho\left(\boldsymbol{r}\right)\left(3r_{i}r_{j}-\delta_{ij}r^{2}\right)d\boldsymbol{r}$,
where $e\rho\left(\boldsymbol{r}\right)$ is the charge distribution
within the molecule.\footnote{When defining the moments it is customary to take the origin in the
center of the positive charge distribution.} These moments are of critical importance as they determine the far-field
potential produced by the molecule , as evident from the monopole
expnasion:\footnote{See reference \citenum{Jackson1999}; we use the Einstein convention
by which repeated Cartesian indices are summed over.} 
\begin{align}
4\pi\epsilon_{0}\varphi\left(\boldsymbol{r}\right) & \equiv e\int\rho\left(\boldsymbol{r}'\right)\left|\boldsymbol{r}-\boldsymbol{r}'\right|^{-1}d\boldsymbol{r}'\label{eq:QMPot}\\
 & =\frac{Q}{r}+\frac{\mu_{i}r_{i}}{r^{3}}+\frac{1}{2}\frac{r_{i}\Theta_{ij}r_{j}}{r^{5}}+....\label{eq:MultipPot}
\end{align}
These low-order ESMs also control the electrostatic interaction energy
$W_{es}$ between the molecule (and through it the forces) with a
weakly non-constant potential $\varphi^{other}\left(\boldsymbol{r}\right)$
resulting from the other molecules or distant charged sources\cite{Jackson1999}:
\begin{equation}
W_{es}=Q\varphi^{other}+\mu_{i}\varphi_{i}^{other}+\frac{1}{6}\Theta_{ij}\varphi_{ij}^{other}+\dots\label{eq:MultipoleEnergy}
\end{equation}
where $\varphi_{i}^{other}=\frac{\partial\varphi^{other}}{\partial r_{i}}$
and $\varphi_{ij}^{other}=\frac{\partial^{2}\varphi^{other}}{\partial r_{i}\partial r_{j}}$
(estimated at a central point within the molecule) etc. This pivotal
\emph{dual }role of ESMs is what drives the requirement that the charge
distribution of the PACs reproduce exactly low-lying molecular ESMs
(MOL-ESMs). This point was discussed at length in ref.~\cite{Verstraelen2016}
where the importance of adherence to the ESMs was demonstrated. An
efficient elegant method for achieving this in as many as possible
moments has been developed \cite{Simmonett2005} although inapplicable
for large molecule charges due to numerical instabilities.\cite{Gilbert2006} 

\begin{figure}
\includegraphics[width=1.1\columnwidth]{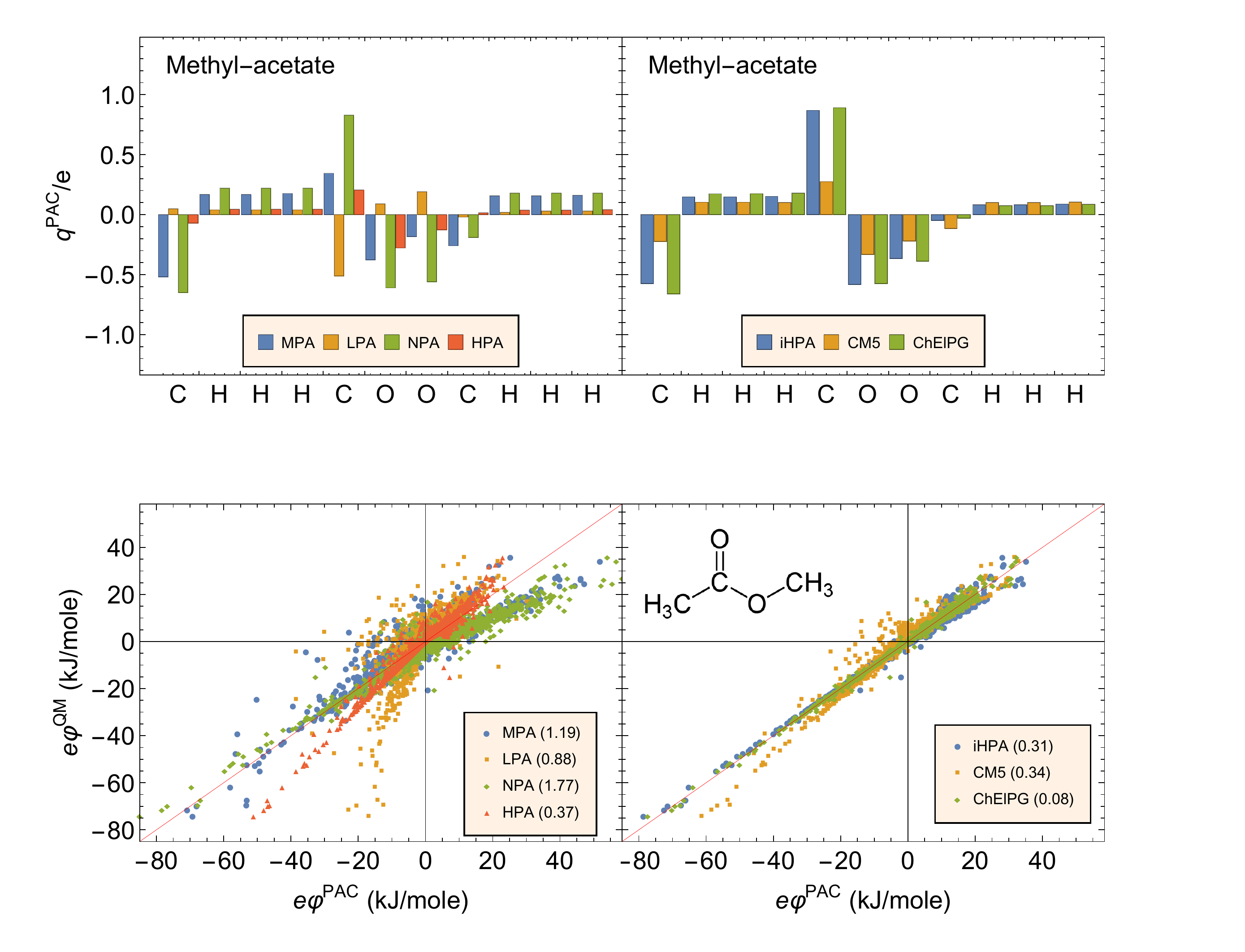}

\caption{\label{fig:MethylAcetate}The PACs (top panels) and the ESP correlation
(bottom panels) for the methyl-acetate molecule, using standard population
analysis methods (left panels) and other PAC methods (right panels).
The mean absolute relative deviation (MARD) of Eq.~\ref{eq:MARD}
appears in parenthesis near each PAC method. Each point in the ESP
correlation plot describes a pair of potentials $\left[\varphi^{PAC}\left(\boldsymbol{r}\right),\varphi^{QM}\left(\boldsymbol{r}\right)\right]$,
the abscissa is the PAC potential (Eq.~\eqref{eq:PotPac}) and the
ordinate the quantum potential (Eq.~\eqref{eq:QMPot}) where $\boldsymbol{r}$
is taken from a subset of grid points of spacing $\Delta x=0.3\mathring{A}$
around the molecule (see description of the grid in Section~\eqref{sec:Results}). }
\end{figure}

Another source of PACs are the quantum mechanical population analysis
(PA) techniques, such as the Mulliken (MPA),\cite{Mulliken1955},
Loewdin (LPA),\cite{Loewdin1950}, Hirshfeld (HPA)\cite{Hirshfeld1977},
and natural population (NPA)\cite{Foster1980} analyses. These PAs
reflect not only the charge distribution but also aspects of the quantum
mechanical wave function. In Fig.~\ref{fig:MethylAcetate} (top left)
we show as bar-plot the PACs produced by these methods applied to
the methyl acetate molecule. It is seen that the different methods
produce sometimes significantly different sets of PACs, even PAC signs
are not preserved! For example, the LPA assigns positive charges to
oxygen atoms, which seems awkward given their high electronegativity.
Furthermore, standard PAs do not reproduce the MOL-ESPs closely, as
shown in the ESP correlation plot of Fig.~\ref{fig:MethylAcetate}
(bottom left), where several PAC-ESPs,
\begin{equation}
\varphi^{PAC}\left(\boldsymbol{r}\right)=\frac{e}{4\pi\epsilon_{0}}\sum_{a}\frac{q_{a}}{\left|\boldsymbol{r}-\boldsymbol{R}_{a}\right|},\label{eq:PotPac}
\end{equation}
are plotted vs. the MOL-ESP $\varphi\left(\boldsymbol{r}\right)$
calculated from the QM density (Eq.~\ref{eq:QMPot}) at a grid point
$\boldsymbol{r}$. The thin red line in the plot corresponds to the
perfectly correlated condition $\varphi^{PAC}=\varphi^{QM}$. In order
to quantify the quality of $\varphi^{PAC}\left(\boldsymbol{r}\right)$
we define the mean absolute relative deviation (MARD) from $\varphi\left(\boldsymbol{r}\right)$
as 
\begin{equation}
MARD\left(\varphi^{PAC},\varphi\right)=\left\langle \left|\frac{\varphi^{PAC}\left(\boldsymbol{r}\right)-\varphi\left(\boldsymbol{r}\right)}{\varphi\left(\boldsymbol{r}\right)}\right|\right\rangle ,\label{eq:MARD}
\end{equation}
where an average is taken over all grid-points $\boldsymbol{r}$ for
which: 1) $\boldsymbol{r}$ is ``outside of the molecule'', i.e.
its distance from any nucleus $a$ is larger than the atomic van-der-Waals
radius $R_{a}^{vdW}$ \cite{Singh1984}) and 2) $\boldsymbol{r}$
is not too far from the molecule, so that its potential $\left|\varphi\left(\boldsymbol{r}\right)\right|$
is not smaller than the threshold value of $e\varphi_{thresh}=0.3eV$.\footnote{Note that the expression in Eq.~\ref{eq:MARD} cannot become singular
due to this requirement.} PACs obtained by ``standard'' PAs have large MARDs: ranging from
0.37 for HPA up to a whopping 1.76 for NPA. On the right panel of
the figure we show data concerning the same molecule, but using the
iterated-Hirshfeld method (iHPA),\cite{Bultinck2007,Bultinck2009,VanDamme2009}
the CM5 method \cite{Marenich2012}, which is a parameterized database
correction to HPA charges, and the ChElPG method\cite{Breneman1990},
which selects PACs that reconstruct the \emph{ab initio }ESP on a
set of grid points as close as possible. The latter approach is taken
here as representative of a class of methods routinely used for PACs
determination. Other members of this method class are the ``charge
from ESPs'' (ChElP)\cite{Chirlian1987}, the Merz-Kollman\cite{Momany1978determination,Singh1984,Besler1990a},
the charge-restraint ESPs (RESP)\cite{Bayly1993,Cornell1993}, atomic
multipoles ESPs\cite{Williams1988}, in combination with molecular
multipoles\cite{Sigfridsson1998} (related to the method proposed
here), the dynamical RESP (D-RESP)\cite{Laio2002a} and Hu-Yang fitting\cite{Hu2007}.The
iHPA, CM5 and ChElPG methods yield much improved description of the
ESP with MARD going from 0.3 for iHPA and CM5 down to 0.08 for ChElPG.
Despite the close ESP fit, ChElPG produces PACs that are usually not
invariant under transformations preserving the point symmetry of the
molecule or under rotations or translations of the nuclei with respect
to the real space grid used to perform the fit. Furthermore, in larger
molecules the PACs of atoms distant from the molecular surface can
become unwieldy large. Both of these issues are discussed in the literature\cite{francl2000pluses,Hu2007}This
instability is likely linked to the fact that the number of parameters
derivable from the ESP in a statistically significant way is considerably
less than the number of atoms.\cite{jakobsen2016searching} Therefore,
iHPA and CM5 are often considered preferred approaches for PACs, although
as seen in the figure, both methods leave ample room for improvement.
Note that the iHPA charges for this molecule are close to the ChElPG
PACs. 

Here, we study a new idea: take PACs which are as close as possible
to a reference set, for example the MPA, HPA or iHPA PACs, but insist
that they reproduce exactly the components of the lowest ESM tensors
(dipole and quadrupole) characterizing the molecular charge distribution.
We formulate a straightforward method to determine such ``minimally-corrected
PACs'' (section~\ref{sec:Method}) and then benchmark the results
using a subset of molecules taken from the database of ref. \cite{Marenich2012}
(section \ref{sec:Results}). Final conclusions are summarized in
section~\ref{sec:Summary-and-conclusions}. All MPA, HPA and iHPA
PACs, as well as the associated MOL-ESPs and MOL-ESMs were computed
using developer versions of Q-Chem 4.3 and 4.4 \cite{Shao2015a} at
the M06-L DFT level \cite{zhao2006new} and using the MG3 semi-diffuse
(MG3S) basis set \cite{Lynch2003}. This functional/basis set combination
was used for developing of the CM5 approach. The CM5, NPA and LPA
results were taken from ref. \citenum{Marenich2012}. 

\section{\label{sec:Method}Method}

Consider a molecule having $A$ nuclei at given Cartesian positions
$\boldsymbol{R}^{a}=\left(R_{x}^{a},\,R_{y}^{a},\,R_{z}^{a}\right)$
($a=1,\dots,A$), for which a QM calculation has determined the charge
density $\rho\left(\boldsymbol{r}\right)$ of the molecule and from
it, low order moments the charge $Q$, the dipole $\mu_{i}$ and the
symmetric traceless quadrupole moment tensor $\Theta_{ij}$. Note
that below, we use the notation $\Theta_{i}^{D}\equiv\Theta_{ii}$
for the diagonal elements of $\Theta$ and $\Theta_{i}^{OD}\equiv\Theta_{jk}$
where $i=x,y,z$ and $ijk$ is a cyclic permutation of $xyz$. For
any set of PACs $\boldsymbol{q}=\left(q_{1},\dots,q_{A}\right)$ we
define the PAC-ESMs as: the monopole (total charge) $Q^{PAC}\equiv e\sum_{a}q_{a}$,
the dipole $\mu_{i}^{PAC}\equiv e\sum_{a}q_{a}R_{i}^{a}$ and the
quadrupole $\Theta_{ij}^{PAC}\equiv e\sum_{a}q_{a}\left(3R_{i}^{a}R_{j}^{a}-\delta_{ij}\left(R^{a}\right)^{2}\right)$,
($i,j=x,y,z$) . Given a set of reference PACs $\boldsymbol{q}^{ref}$
we seek to determine a ``minimally-corrected'' set of PACs $\boldsymbol{q}^{mc}=\boldsymbol{q}^{ref}+\Delta\boldsymbol{q}$
such that that the size of the correction $\left\Vert \Delta\boldsymbol{q}\right\Vert ^{2}=\Delta\boldsymbol{q}\cdot\Delta\boldsymbol{q}$
is \emph{minimal }but the multipoles are equal to the QM determined
multipoles, i.e. the following constraints are satisfied: 
\begin{align}
c & =Q-Q^{PAC}=0,\nonumber \\
c_{i} & =\mu_{i}-\mu_{i}^{PAC}=0,\label{eq:Constraints}\\
c_{ij} & =\Theta_{ij}-\Theta_{ij}^{PAC}=0.\nonumber 
\end{align}
Note that the number of constraints (denoted $C$) in Eq.~\eqref{eq:Constraints}
is $9$ and not $13$ since the electric quadrupole tensor is symmetric
and traceless. Point symmetries can reduce this number of constraints
further. If, for example, both positive and negative charge densities
are symmetric against the reflection through a plane (the x-y plane,
for example) then there are 3 constraint less (one from the z component
of the dipole and and 2 from XZ and YZ components of the quadrupole,
which are zero by symmetry). Only when the number of atoms $A$ in
the molecule is greater than the number of constraints $C$ can we
hope to reproduce the constraints exactly. We therefore demand that
$A>9$ and use the $A-C$ additional ``degrees of freedom'' to minimize
the deviance $\Delta\boldsymbol{q}$. When $2\le A\le9$ we avoid
the quadrupole moment constraint and use only the dipole moment constraint.

We are led to consider the Lagrangian
\begin{align}
L_{mcDQ} & =\frac{1}{2}\sum_{a}\left(q_{a}^{mc}-q_{a}^{ref}\right)^{2}-\lambda c-\lambda_{i}c_{i}\label{eq:Lagrangian}\\
 & -\sum_{xy,yz,zx}\lambda_{ij}^{OD}c_{ij}-\sum_{x,y,z}\lambda_{ii}^{D}c_{ii}\nonumber 
\end{align}
as a function of the $A$ $q's$ and the ten Lagrange multipliers:
one $\lambda,$ three $\lambda_{i}$'s , three diagonal $\lambda_{ii}^{D}$
and three off-diagonal$\lambda_{jk}^{OD}$ where $i=x,y,z$ and $ijk$
is a cyclic permutation of $xyz$. Taking derivatives with respect
to these variables and equating to zero leads to the following set
of $\left(10+A\right)$ linear equations in $\left(10+A\right)$ unknowns,
given here in block-matrix/vector form\footnote{Since the matrix is dominated by zero's one can formulate the linear
equation in a more concise way. However, this form is straightforward
to derive and manipulate when there are instabilities, discussed later.}:

\begin{align}
S & \left(\begin{array}{c}
q_{1}^{mc}\\
\vdots\\
q_{A}^{mc}\\
\lambda_{1\times1}\\
\lambda_{3\times1}\\
\lambda_{3\times1}^{D}\\
\lambda_{3\times1}^{OD}
\end{array}\right)=\left(\begin{array}{c}
q_{1}^{ref}\\
\vdots\\
q_{A}^{ref}\\
Q_{1\times1}\\
\mu_{3\times1}\\
\Theta_{3\times1}^{D}\\
\Theta_{3\times1}^{OD}
\end{array}\right).\label{eq:Ax=00003Db}
\end{align}
The $\left(A+10\right)\times\left(A+10\right)$ matrix $S$ is of
the following form:

{\scriptsize{}
\begin{equation}
S=\left(\begin{array}{ccccccccccccc}
\text{\textSFi} & \cdots & \text{\textSFiii} & -1 & \text{\textSFi} &  & \text{\textSFiii} & \text{\textSFi} &  & \text{\textSFiii} & \text{\textSFi} &  & \text{\textSFiii}\\
\vdots & I_{A\times A} & \vdots & \vdots & \vdots & -D_{A\times3} & \vdots & \vdots & -T_{A\times3}^{D} & \vdots & \vdots & -T_{A\times3}^{OD} & \vdots\\
\text{\textSFii} & \cdots & \text{\textSFiv} & -1 & \text{\textSFii} &  & \text{\textSFiv} & \text{\textSFii} &  & \text{\textSFiv} & \text{\textSFii} &  & \text{\textSFiv}\\
1 & \cdots & 1 & 0 & 0 & 0 & 0 & 0 & 0 & 0 & 0 & 0 & 0\\
\text{\textSFi} & \cdots & \text{\textSFiii} & 0 & 0 & 0 & 0 & 0 & 0 & 0 & 0 & 0 & 0\\
 & D_{3\times A} &  & 0 & 0 & 0 & 0 & 0 & 0 & 0 & 0 & 0 & 0\\
\text{\textSFii} & \cdots & \text{\textSFiv} & 0 & 0 & 0 & 0 & 0 & 0 & 0 & 0 & 0 & 0\\
\text{\textSFi} & \cdots & \text{\textSFiii} & 0 & 0 & 0 & 0 & 0 & 0 & 0 & 0 & 0 & 0\\
 & T_{3\times A}^{D} &  & 0 & 0 & 0 & 0 & 0 & 0 & 0 & 0 & 0 & 0\\
\text{\textSFii} & \cdots & \text{\textSFiv} & 0 & 0 & 0 & 0 & 0 & 0 & 0 & 0 & 0 & 0\\
\text{\textSFi} & \cdots & \text{\textSFiii} & 0 & 0 & 0 & 0 & 0 & 0 & 0 & 0 & 0 & 0\\
 & T_{3\times A}^{OD} &  & 0 & 0 & 0 & 0 & 0 & 0 & 0 & 0 & 0 & 0\\
\text{\textSFii} & \cdots & \text{\textSFiv} & 0 & 0 & 0 & 0 & 0 & 0 & 0 & 0 & 0 & 0
\end{array}\right)\label{eq:Smat}
\end{equation}
}and depends \emph{only }on the location of the atomic nuclei. The
matrix is composed of blocks: the $I_{A\times A}$ block is a $A\times A$
unit matrix, $D_{A\times3}$, $T_{A\times3}^{D}$ and $T_{A\times3}^{OD}$
are matrices of dimension $A\times3$ ($3$ columns each of length
$A$) of matrix elements: $D_{ai}=eR_{i}^{a}$ and $T_{ai}^{D}=e\left(3R_{i}^{a}R_{i}^{a}-\Vert R^{a}\Vert^{2}\right)$
for $a=1,\dots,A$ and $i=x,y,z$ and and $T_{ai}^{OD}=3eR_{j}^{a}R_{k}^{a}$
(where the ordered set $i,j,k$ is a cyclic permutation of $x,y,z$).
The $D_{3\times A}$, $T_{3\times A}^{D}$, and $T_{3\times A}^{OD}$
blocks are respectively the transposed matrices. The $A+10$ column-vector
on the left-hand-side of Eq.~\ref{eq:Ax=00003Db} includes the unknowns,
the $A$ partial charges $q_{a}^{mc}$ and the $\lambda's$, the ten
Lagrange multipliers for the ten constraints. The $10+A$ column-vector
on the right-hand-side has $A$ values of the reference charges $q_{a}^{ref}$,
followed by the total charge on the molecule $Q$, then the three
values of the QM dipole moment $\mu_{i}$ followed by the three values
of the diagonal elements of the given QM quadrupole tensor $\Theta_{i}^{D}=\Theta_{ii}$
and finally the QM values of the three off-diagonal elements $\Theta_{i}^{OD}=\Theta_{jk}$
where $i=x,y,z$ and $ijk$ is a cyclic permutation of $xyz$). A
similar equation holds for the mcD method, where the six last rows
are erased from $S$ and from the column vectors and the six right
columns are erased from $S$ as well. This leaves us with a $\left(A+4\right)\times\left(A+4\right)$
system of equations. 

The structural matrix $S$ may become singular or rank deficient.
One trivial source for singularity the use of 3 diagonal constraints
while their sum is composed to be zero. The use of the singular-value-decomposition
pseudo-inverse \cite{Golub1996} for solving Eq.~\ref{eq:Ax=00003Db}
helps to bypass such a singularity. A more delicate source of singularities
may arise from symmetry. For example, when the molecule is perfectly
planar (or has a plane of symmetry) in the x-y plane then the row
corresponding to the dipole in the z direction $D_{az}=eR_{z}^{a}$
must be identically zero and the matrix $S$ will be rank deficient.
In this case the $T_{ai}^{OD}$ with $i=x$and $y$ must also be zero).
In these cases the SVD pseudoinverse will automatically eliminate
constraints that cannot be met due to this kind of symmetry. But for
near-symmetrical configurations, instabilities may exist. In cases
such as these we can still spot problems by examining the values of
the Lagrange multipliers $\lambda,$$\lambda_{i}$ and $\lambda_{ij}$in
the solution vector of Eq.~\ref{eq:Ax=00003Db}. The Lagrange multiplier
is equal to the derivative of the minimal value of the Lagrangian
$L$ with respect to the constraint value ($Q$, $\mu_{i}$ and $\Theta_{ij}$,
respectively). Thus if the \emph{ab initio }dipole moment $\mu_{x}$
is given to precision $\delta\mu_{x}$, the product $\left|\lambda_{x}\delta\mu_{x}\right|$is
expected to be the error in the minimal value of $L$. Clearly, the
minimizing procedure is meaningless unless this error is much smaller
than 1. Hence, it is important to eliminate ``offending'' constraints
from the matrix equation (the corresponding row and column in the
matrix and the entry in the column vectors) for those having large
Lagrange multipliers. We know that \emph{ab initio }multipole properties
are usually given to 3 digits hence we eliminate constraints corresponding
to Lagrange multipliers large than 1000. The reduced equation is then
solved and the remaining Lagrange multipliers are examined again.
We repeat such elimination until all Lagrange multipliers have proper
magnitudes. This pruning procedure helps avoid cases where small inaccuracies
of the input data dominate the final result. Within the molecules
studied here such a pruning procedure was used only for few cases
of molecules having a near plane symmetry.

When symmetry is active, our procedures reduce the number of constraints
and hence the number of \emph{independent }$q_{a}$'s (called number
of degrees of freedom (NDOFs)). For example, the water molecule has
3 nuclei but due to symmetry the two H nuclei will have the same PACs
and so NDOF=2. Due to the symmetry only the dipole moment in the direction
of the $C_{2}$ axis is a constraint (the components perpendicular
to the C2 axis are zero by symmetry). Together with the charge of
water (0) we already have 2 constraints so one must give up imposing
the quadrupole moment for water. 

\begin{figure}[h]
\includegraphics[width=1\columnwidth]{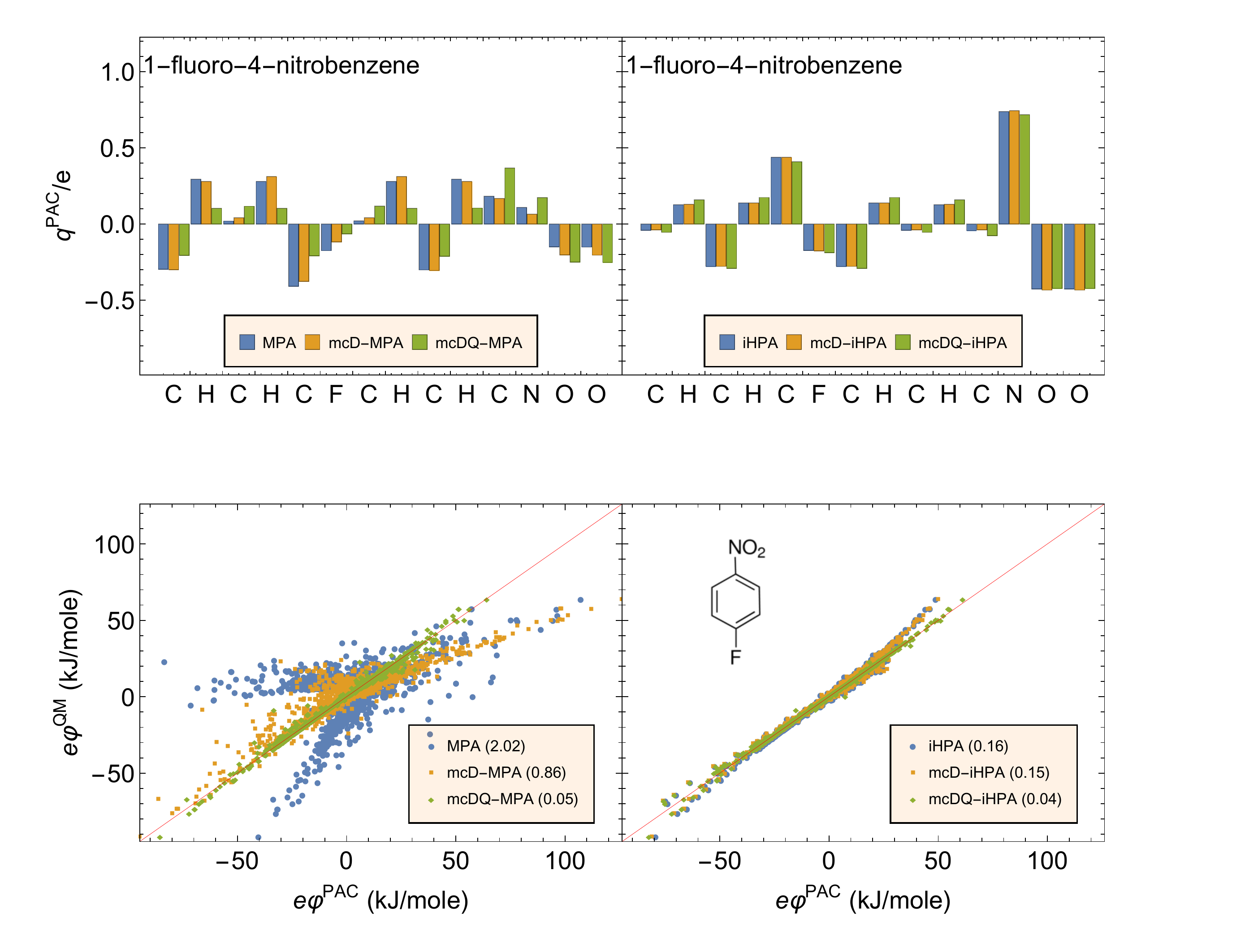}

\caption{\label{fig:Efficacy}The PACs (top panels) and and ESP correlation
plots (bottom panels) for MPA, mcD-MPA and mcDQ-MPA (left panels)
and iHPA, mcD-iHPA and mcDQ-iHPA (right panels) in the 1-fluoro-4-nitrobenzene
molecule. The mean absolute relative deviations (MARD) appears in
parenthesis near each PAC method. }
\end{figure}

\section{\label{sec:Results}Results }

To demonstrate the efficacy of the method we show in Fig.~\ref{fig:Efficacy}
the MPA and iHPA PACs and their ESP correlation plots before and after
applying the minimal corrections required for imposing dipole and
quadrupole moments (denoted mcD/mcDQ-MPA and mcD/mcDQ-iHPA respectively).\footnote{Minimally-corrected PACs that reproduce only the dipole moment are
designated mcD and those that reproduce the components of the dipole
and the quadrupole ESMs are designated mcDQ.} 

\begin{figure}
\includegraphics[width=1\textwidth]{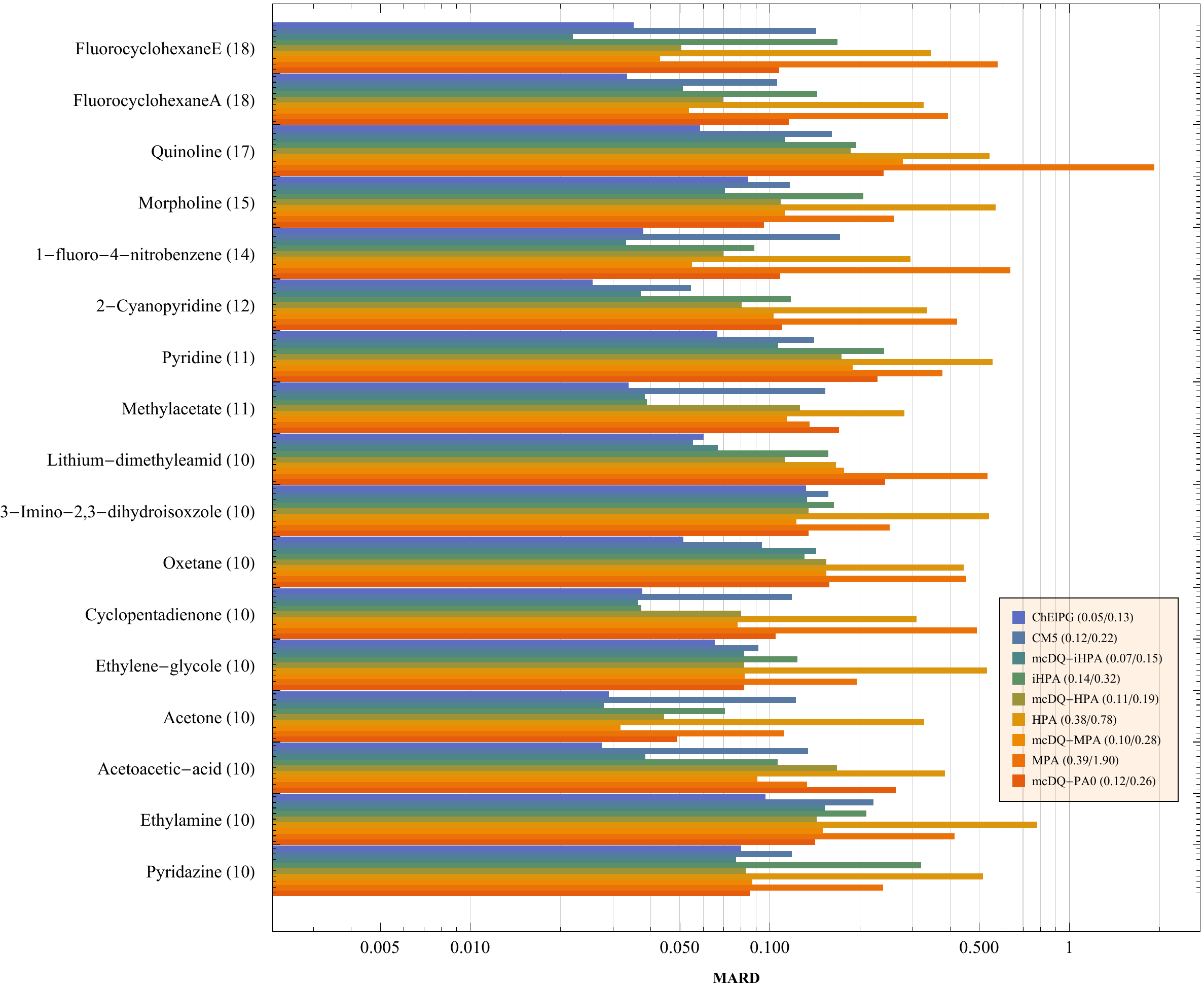}

\caption{\label{fig:HPA-Histog}The MARD (Eq.~\eqref{eq:MARD}) of various
PACs schemes for a subset of molecule taken from ref. \cite{Marenich2012}.
Numbers in parenthesis appearing near the molecule names indicate
the number of atoms in that molecule. The pair of numbers (median/max)
appearing in the legend box near each scheme is, respectively, the
median and maximum of the relative deviance taken over the shown set
molecules.}
\end{figure}

Notice that the MPA-ESP has low correlation with the MOL-ESP, as can
be evident visually and also by the reported MARD of 2. The mcD corrections
improve the ESP but only mcDQ corrections show high quality ESP (with
MARD of 0.05). In accordance with previous reports,\cite{VanDamme2009}
the iHPA ESP already correlates nicely with the MOL-ESP (MARD of 0.16)
but the mcDQ-iHPA improves the correlation significantly and the MARD
reduces by a factor of 4. For this molecule, both mcDQ-MPA and mcDQ-iHPA
have similar MARDs but this is not typical, for most molecules the
mcDQ-iHPA MARDs are much smaller than those of mcDQ-MPA (see Fig.\ref{fig:HPA-Histog}).
The mcDQ-MPA PACs are not drastically different from the MPA PACs
yet their MARDs are considerably lower. This shows the power of the
minimally-corrected PACs, where a small change in PACs can improve
the PAC based ESP considerably. 

In Fig.~\ref{fig:HPA-Histog} we display a log-scale bar-plot of
MARDs of several PAC-based potentials on selected molecules containing
10-18 atoms. Each PAC method can be characterized by a pair of numbers
(shown in parenthesis within the legend box) indicating the median/maximal
MARD taken over the given set of molecules. The PACs obtained by minimally-correcting
the $q\equiv0$ reference (called 0PA) are actually the minimal PACs
that give the dipole and quadrupole of the molecules. It is seen that
their correlation with the exact ESP is considerably higher than that
of MPA and HPA, somewhat similar to that of mcDQ-MPA and mcDQ-HPA,
close to that of CM5. This goes to show that the fit of just the dipole
and quadrupole, keeping the charges as small as possible gives a reasonably
behaved ESP, although in general, for very large molecules the mcDQ-0PA
performance may degrade with size compared to the PA methods. We see
that MPA and HPA have similar MARDs while iHPA seems to give considerably
smaller MARDs (by a factor 2-3). The minimal corrected (mcDQ) to MPA
and HPA yield smaller MARDs by a factor 4 and for iHPA by a factor
2. Altogether the mcDQ significantly improves the ESP. The mcDQ-iHPA
median MARD is 7\% is similar to that of ChElPG (5\%). 

It is worthwhile to examine the sensitivity of the MARD estimation
with respect to the distance of grid points from the nearest nuclei.
In Fig.~\ref{fig:HPA-Histog} all sampling grid points were at a
distance larger than $1.5\times r_{vdW}$ from any atom. When MARD
is estimated using points further way (distance larger than a value
of $2\times r_{vdW}$) the iHPA MARD dropped from 0.14 to 0.09 and
mcDQ-iHPA MARD dropped from 0.07 to 0.03. ChElPG MARD also reduced,
from 0.05 to 0.03. This finding is consistent with the fact that the
MCDQ methods provide an asymptotically exact far-field ESP resulting
from their reconstruction of the molecular dipole and quadrupole moments.

In Table~\ref{tab:dq} we show, for each set of PACs the magnitude
of the charge correction $\left\Vert \Delta q\right\Vert _{\infty}$.
For a given molecule the mcD correction is largest for 0PA and then
for MPA and HPA and it is smallest for iHPA. mcDQ corrections are
in general considerably larger than mcD but in both methods $\left\Vert \Delta q\right\Vert _{\infty}$
decreases as the number of atoms in the molecule grows. This is due
to the fact that in large systems even small charge shifts have a
large affect on the dipole and the quadrupole moments.

\begin{table}
{\scriptsize{}}%
\begin{tabular}{|l|l|l|l|l|r@{\extracolsep{0pt}.}l|r@{\extracolsep{0pt}.}l|r@{\extracolsep{0pt}.}l|r@{\extracolsep{0pt}.}l|r@{\extracolsep{0pt}.}l|r@{\extracolsep{0pt}.}l|r@{\extracolsep{0pt}.}l|r@{\extracolsep{0pt}.}l|}
\hline 
\multirow{2}{*}{{\scriptsize{}Molecule}} & \multirow{2}{*}{{\scriptsize{}Sym}} & \multirow{2}{*}{{\scriptsize{}$A$}} & \multirow{2}{*}{{\scriptsize{}$F$}} & \multirow{2}{*}{{\scriptsize{}$C$}} & \multicolumn{8}{c|}{{\scriptsize{}$\left\Vert \Delta q\right\Vert _{\infty}$ (mcD)}} & \multicolumn{8}{c|}{{\scriptsize{}$\left\Vert \Delta q\right\Vert _{\infty}$ (mcDQ)}}\tabularnewline
\cline{6-21} 
 &  &  &  &  & \multicolumn{2}{c|}{{\scriptsize{}0PA}} & \multicolumn{2}{c|}{{\scriptsize{}MPA}} & \multicolumn{2}{c|}{{\scriptsize{}HPA}} & \multicolumn{2}{c|}{{\scriptsize{}iHPA}} & \multicolumn{2}{c|}{{\scriptsize{}0PA}} & \multicolumn{2}{c|}{{\scriptsize{}MPA}} & \multicolumn{2}{c|}{{\scriptsize{}HPA}} & \multicolumn{2}{c|}{{\scriptsize{}iHPA}}\tabularnewline
\hline 
\hline 
{\scriptsize{}Pyridazine} & {\scriptsize{}$C_{2v}$} & {\scriptsize{}10} & {\scriptsize{}7} & {\scriptsize{}6} & {\scriptsize{}0}&{\scriptsize{}11} & {\scriptsize{}0}&{\scriptsize{}02} & {\scriptsize{}0}&{\scriptsize{}06} & {\scriptsize{}0}&{\scriptsize{}04} & {\scriptsize{}0}&{\scriptsize{}27} & {\scriptsize{}0}&{\scriptsize{}29} & {\scriptsize{}0}&{\scriptsize{}17} & {\scriptsize{}0}&{\scriptsize{}12}\tabularnewline
\hline 
{\scriptsize{}Ethylamine} & {\scriptsize{}$C_{s}$} & {\scriptsize{}10} & {\scriptsize{}7} & {\scriptsize{}6} & {\scriptsize{}0}&{\scriptsize{}04} & {\scriptsize{}0}&{\scriptsize{}01} & {\scriptsize{}0}&{\scriptsize{}03} & {\scriptsize{}0}&{\scriptsize{}01} & {\scriptsize{}1}&{\scriptsize{}17} & {\scriptsize{}1}&{\scriptsize{}61} & {\scriptsize{}1}&{\scriptsize{}22} & {\scriptsize{}1}&{\scriptsize{}28}\tabularnewline
\hline 
{\scriptsize{}Acetoacetic acid} & {\scriptsize{}$C_{1}$} & {\scriptsize{}10} & {\scriptsize{}9} & {\scriptsize{}6} & {\scriptsize{}0}&{\scriptsize{}07} & {\scriptsize{}0}&{\scriptsize{}02} & {\scriptsize{}0}&{\scriptsize{}03} & {\scriptsize{}0}&{\scriptsize{}01} & {\scriptsize{}0}&{\scriptsize{}54} & {\scriptsize{}0}&{\scriptsize{}26} & {\scriptsize{}0}&{\scriptsize{}25} & {\scriptsize{}0}&{\scriptsize{}21}\tabularnewline
\hline 
{\scriptsize{}Acetone} & {\scriptsize{}$C_{2v}$} & {\scriptsize{}10} & {\scriptsize{}5} & {\scriptsize{}5} & {\scriptsize{}0}&{\scriptsize{}17} & {\scriptsize{}0}&{\scriptsize{}02} & {\scriptsize{}0}&{\scriptsize{}05} & {\scriptsize{}0}&{\scriptsize{}01} & {\scriptsize{}0}&{\scriptsize{}52} & {\scriptsize{}0}&{\scriptsize{}48} & {\scriptsize{}0}&{\scriptsize{}34} & {\scriptsize{}0}&{\scriptsize{}08}\tabularnewline
\hline 
{\scriptsize{}Ethylene-glycol} & {\scriptsize{}$C_{1}$} & {\scriptsize{}10} & {\scriptsize{}10} & \textbf{\scriptsize{}9} & {\scriptsize{}0}&{\scriptsize{}09} & {\scriptsize{}0}&{\scriptsize{}03} & {\scriptsize{}0}&{\scriptsize{}05} & {\scriptsize{}0}&{\scriptsize{}02} & {\scriptsize{}0}&{\scriptsize{}67} & {\scriptsize{}0}&{\scriptsize{}88} & {\scriptsize{}0}&{\scriptsize{}64} & {\scriptsize{}0}&{\scriptsize{}54}\tabularnewline
\hline 
{\scriptsize{}Cyclopentadienone} & {\scriptsize{}$C_{2v}$} & {\scriptsize{}10} & {\scriptsize{}6} & \textbf{\scriptsize{}6} & {\scriptsize{}0}&{\scriptsize{}09} & {\scriptsize{}0}&{\scriptsize{}04} & {\scriptsize{}0}&{\scriptsize{}03} & {\scriptsize{}0}&{\scriptsize{}00} & {\scriptsize{}0}&{\scriptsize{}19} & {\scriptsize{}0}&{\scriptsize{}22} & {\scriptsize{}0}&{\scriptsize{}05} & {\scriptsize{}0}&{\scriptsize{}02}\tabularnewline
\hline 
{\scriptsize{}Oxetane} & {\scriptsize{}$C_{s}$} & {\scriptsize{}10} & {\scriptsize{}7} & {\scriptsize{}6} & {\scriptsize{}0}&{\scriptsize{}08} & {\scriptsize{}0}&{\scriptsize{}05} & {\scriptsize{}0}&{\scriptsize{}03} & {\scriptsize{}0}&{\scriptsize{}01} & {\scriptsize{}1}&{\scriptsize{}09} & {\scriptsize{}1}&{\scriptsize{}29} & {\scriptsize{}1}&{\scriptsize{}09} & {\scriptsize{}1}&{\scriptsize{}09}\tabularnewline
\hline 
{\scriptsize{}3-Imino-2,3-dihydroisoxzole} & {\scriptsize{}$C_{s}$} & {\scriptsize{}10} & {\scriptsize{}8} & \textbf{\scriptsize{}6} & {\scriptsize{}0}&{\scriptsize{}05} & {\scriptsize{}0}&{\scriptsize{}02} & {\scriptsize{}0}&{\scriptsize{}02} & {\scriptsize{}0}&{\scriptsize{}01} & {\scriptsize{}0}&{\scriptsize{}79} & {\scriptsize{}0}&{\scriptsize{}83} & {\scriptsize{}0}&{\scriptsize{}69} & {\scriptsize{}0}&{\scriptsize{}41}\tabularnewline
\hline 
{\scriptsize{}Lithium-dimethylamine} & {\scriptsize{}$C_{2v}$} & {\scriptsize{}10} & {\scriptsize{}6} & \textbf{\scriptsize{}4} & {\scriptsize{}0}&{\scriptsize{}31} & {\scriptsize{}0}&{\scriptsize{}16} & {\scriptsize{}0}&{\scriptsize{}05} & {\scriptsize{}0}&{\scriptsize{}04} & {\scriptsize{}0}&{\scriptsize{}31} & {\scriptsize{}0}&{\scriptsize{}16} & {\scriptsize{}0}&{\scriptsize{}05} & {\scriptsize{}0}&{\scriptsize{}04}\tabularnewline
\hline 
{\scriptsize{}Methylacetate} & {\scriptsize{}$C_{s}$} & {\scriptsize{}11} & {\scriptsize{}9} & \textbf{\scriptsize{}6} & {\scriptsize{}0}&{\scriptsize{}10} & {\scriptsize{}0}&{\scriptsize{}02} & {\scriptsize{}0}&{\scriptsize{}02} & {\scriptsize{}0}&{\scriptsize{}01} & {\scriptsize{}0}&{\scriptsize{}53} & {\scriptsize{}0}&{\scriptsize{}22} & {\scriptsize{}0}&{\scriptsize{}26} & {\scriptsize{}0}&{\scriptsize{}01}\tabularnewline
\hline 
{\scriptsize{}Pyridine} & {\scriptsize{}$C_{2v}$} & {\scriptsize{}11} & {\scriptsize{}7} & \textbf{\scriptsize{}4} & {\scriptsize{}0}&{\scriptsize{}06} & {\scriptsize{}0}&{\scriptsize{}02} & {\scriptsize{}0}&{\scriptsize{}03} & {\scriptsize{}0}&{\scriptsize{}01} & {\scriptsize{}0}&{\scriptsize{}15} & {\scriptsize{}0}&{\scriptsize{}23} & {\scriptsize{}0}&{\scriptsize{}09} & {\scriptsize{}0}&{\scriptsize{}03}\tabularnewline
\hline 
{\scriptsize{}2-Cyanopyridine} & {\scriptsize{}$C_{s}$} & {\scriptsize{}12} & {\scriptsize{}12} & \textbf{\scriptsize{}6} & {\scriptsize{}0}&{\scriptsize{}12} & {\scriptsize{}0}&{\scriptsize{}05} & {\scriptsize{}0}&{\scriptsize{}04} & {\scriptsize{}0}&{\scriptsize{}01} & {\scriptsize{}0}&{\scriptsize{}19} & {\scriptsize{}0}&{\scriptsize{}19} & {\scriptsize{}0}&{\scriptsize{}07} & {\scriptsize{}0}&{\scriptsize{}03}\tabularnewline
\hline 
{\scriptsize{}1-fluoro-4-nitrobenzene} & {\scriptsize{}$C_{2v}$} & {\scriptsize{}14} & {\scriptsize{}9} & \textbf{\scriptsize{}6{*}} & {\scriptsize{}0}&{\scriptsize{}05} & {\scriptsize{}0}&{\scriptsize{}06} & {\scriptsize{}0}&{\scriptsize{}01} & {\scriptsize{}0}&{\scriptsize{}01} & {\scriptsize{}0}&{\scriptsize{}13} & {\scriptsize{}0}&{\scriptsize{}20} & {\scriptsize{}0}&{\scriptsize{}05} & {\scriptsize{}0}&{\scriptsize{}04}\tabularnewline
\hline 
{\scriptsize{}Morpholine} & {\scriptsize{}$C_{s}$} & {\scriptsize{}15} & {\scriptsize{}9} & \textbf{\scriptsize{}6} & {\scriptsize{}0}&{\scriptsize{}03} & {\scriptsize{}0}&{\scriptsize{}01} & {\scriptsize{}0}&{\scriptsize{}02} & {\scriptsize{}0}&{\scriptsize{}01} & {\scriptsize{}0}&{\scriptsize{}47} & {\scriptsize{}0}&{\scriptsize{}04} & {\scriptsize{}0}&{\scriptsize{}33} & {\scriptsize{}0}&{\scriptsize{}08}\tabularnewline
\hline 
{\scriptsize{}Quinoline} & {\scriptsize{}$C_{s}$} & {\scriptsize{}17} & {\scriptsize{}17} & \textbf{\scriptsize{}6} & {\scriptsize{}0}&{\scriptsize{}03} & {\scriptsize{}0}&{\scriptsize{}12} & {\scriptsize{}0}&{\scriptsize{}02} & {\scriptsize{}0}&{\scriptsize{}01} & {\scriptsize{}0}&{\scriptsize{}10} & {\scriptsize{}0}&{\scriptsize{}26} & {\scriptsize{}0}&{\scriptsize{}05} & {\scriptsize{}0}&{\scriptsize{}02}\tabularnewline
\hline 
{\scriptsize{}Fluorocyclohexane (A)} & {\scriptsize{}$C_{s}$} & {\scriptsize{}18} & {\scriptsize{}12} & \textbf{\scriptsize{}6} & {\scriptsize{}0}&{\scriptsize{}05} & {\scriptsize{}0}&{\scriptsize{}03} & {\scriptsize{}0}&{\scriptsize{}02} & {\scriptsize{}0}&{\scriptsize{}01} & {\scriptsize{}0}&{\scriptsize{}17} & {\scriptsize{}0}&{\scriptsize{}06} & {\scriptsize{}0}&{\scriptsize{}05} & {\scriptsize{}0}&{\scriptsize{}02}\tabularnewline
\hline 
{\scriptsize{}Fluorocyclohexane (E)} & {\scriptsize{}$C_{s}$} & {\scriptsize{}18} & {\scriptsize{}12} & \textbf{\scriptsize{}6} & {\scriptsize{}0}&{\scriptsize{}04} & {\scriptsize{}0}&{\scriptsize{}03} & {\scriptsize{}0}&{\scriptsize{}01} & {\scriptsize{}0}&{\scriptsize{}01} & {\scriptsize{}0}&{\scriptsize{}17} & {\scriptsize{}0}&{\scriptsize{}08} & {\scriptsize{}0}&{\scriptsize{}07} & {\scriptsize{}0}&{\scriptsize{}04}\tabularnewline
\hline 
\multicolumn{5}{|l|}{\textbf{\scriptsize{}Median}} & \textbf{\scriptsize{}0}&\textbf{\scriptsize{}07} & \textbf{\scriptsize{}0}&\textbf{\scriptsize{}03} & \textbf{\scriptsize{}0}&\textbf{\scriptsize{}03} & \textbf{\scriptsize{}0}&\textbf{\scriptsize{}01} & \textbf{\scriptsize{}0}&\textbf{\scriptsize{}31} & \textbf{\scriptsize{}0}&\textbf{\scriptsize{}23} & \textbf{\scriptsize{}0}&\textbf{\scriptsize{}17} & \textbf{\scriptsize{}0}&\textbf{\scriptsize{}04}\tabularnewline
\hline 
\multicolumn{5}{|l|}{\textbf{\scriptsize{}Max}} & \textbf{\scriptsize{}0}&\textbf{\scriptsize{}31} & \textbf{\scriptsize{}0}&\textbf{\scriptsize{}16} & \textbf{\scriptsize{}0}&\textbf{\scriptsize{}06} & \textbf{\scriptsize{}0}&\textbf{\scriptsize{}04} & \textbf{\scriptsize{}1}&\textbf{\scriptsize{}17} & \textbf{\scriptsize{}1}&\textbf{\scriptsize{}61} & \textbf{\scriptsize{}1}&\textbf{\scriptsize{}22} & \textbf{\scriptsize{}1}&\textbf{\scriptsize{}28}\tabularnewline
\hline 
\end{tabular}{\scriptsize \par}

\caption{\label{tab:dq}The PAC change $\left\Vert \Delta q\right\Vert _{\infty}=max_{1\le a\le A}\left|\Delta q_{a}\right|$
induced by mcD and mcDQ for 0PA (where the reference PACs are all
zero), MPA, HPA and iHPA for the set of molecules of Fig.~\ref{fig:HPA-Histog}.
Also shown the number of atoms $A$ the number of degrees of freedom
$F$ and the number of constraints $C$ for each molecule.}
\end{table}

In table~\ref{tab:Bitzuim} we summarize the MARD statistics (median
and maximal) for for four sets of reference charges: 0PA (reference
charges are equal to zero) and MPA, HPA, iHPA. The efficiency of the
mc procedure is apparent for MPA, HPA and iHPA, where the mcD reduces
the median/maximal MARD by about a factor of 2. mcDQ reduces the MARD
further, by a factor of 3 for 0PA and \textasciitilde{}2 for MPA and
HPA and only 1.1 for iHPA. We thus see that iHPA reconstruction of
the ESP strongly benefits from a dipole correction and, interestingly,
much less a quadrupole correction.

\begin{table}
\begin{tabular}{|c|c|c|c|c|}
\hline 
 & 0PA & MPA & HPA & iHPA\tabularnewline
\hline 
\hline 
no-correction & NA & 0.39/1.90 & 0.38/0.78 & 0.14/0.32\tabularnewline
\hline 
mcD & 0.41/0.67 & 0.18/0.60 & 0.18/0.53 & 0.08/0.17\tabularnewline
\hline 
mcDQ & 0.12/0.26 & 0.10/0.28 & 0.11/0.19 & 0.07/0.15\tabularnewline
\hline 
\end{tabular}

\caption{\label{tab:Bitzuim}The median/maximal MARD (for the set of molecules
used above) determined for each PAC reference found for: non-corrected,
and minimally-corrected schemes, mcD and mcDQ . }
\end{table}

\begin{figure}
\includegraphics[width=0.7\columnwidth]{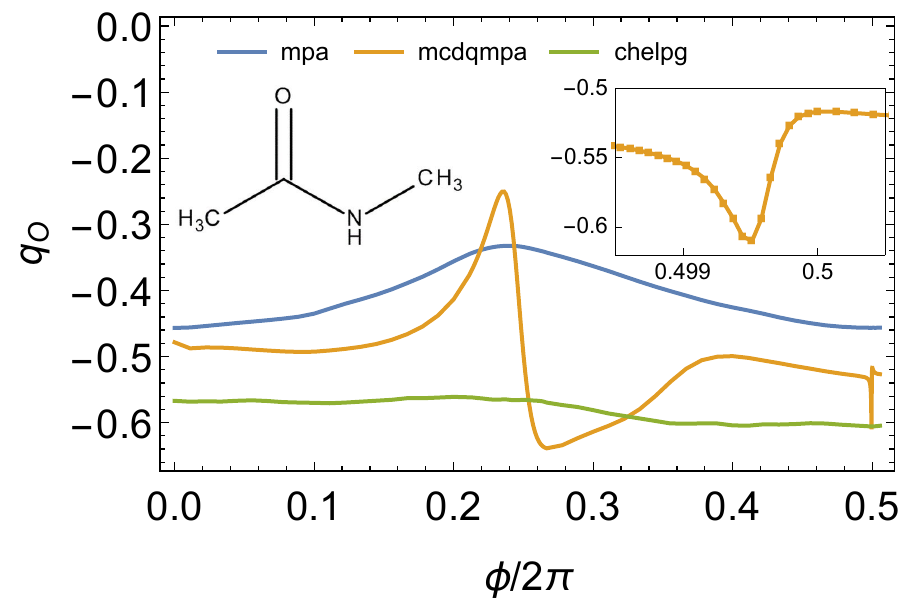}

\caption{\label{fig:q_O(phi)}The partial charge, determined by MPA, mcDQ-MPA
and ChElPG on the oxygen atom as a function of the O-C-N-H dihedral
angle $\phi$ in N-methylethanamide (NMA). Inset om the right shows
the sharp feature near $\phi=\pi$. }
\end{figure}

PACs are sometimes used when molecules distort. In this case, it is
important that the they remain continuous under the distortion, so
as to enable force calculations. The MPA/HPA/iHPA do not show non-smooth
behavior and the mcDQ which is a minimization procedure does not show
it as well.\footnote{We cannot rule out possible issues if the matrix $S$ of Eq.~\ref{eq:Smat}
becomes rank deficient. However, we believe this is an unlikely or
quite rare event.} In \ref{fig:q_O(phi)} we show the MPA, mcDQ-MPA and ChElPG PACs
of the oxygen atom in N-methylethanamide\cite{Hu2007} the as a function
of the dihedral angle $\phi$. It is seen that as the angle increases
from 0 the mcDQ-MPA PAC slightly decreases and then increase rapidly
followed by a rapid yet continuous drop near $\phi_{c}=0.25$ from
a value of $q_{O}=-0.25$ to $q_{O}\approx-0.64$. An additional very
sharp feature is seen near $\phi=\pi$. We have checked that this
sharp feature is not discontinuous (see inset in Fig.~\ref{fig:q_O(phi)})
and that the matrix $S$ of Eq.~\ref{eq:Smat} does not become rank
deficient. Similar behavior is seen for the PACs of other atoms. We
thus conclude that the charges change continuously although sometimes
very rapid charge fluctuations can occur. 

\section{\label{sec:Summary-and-conclusions}Summary and conclusions}

We have studied a new scheme for minimally correcting reference PACs
so that they reproduce the exact dipole and quadrupole moments of
a molecule and we found that such a minimal correction greatly improves
the correlation of the PAC-ESP with respect to the MOL-ESP. The minimal
correction scheme does not alter symmetry properties of the reference
PACs. Hence, minimally-corrected PACs (mc-PACs) based on MPA, HPA,
iHPA fully respect the point-symmetry and rotational/translational
symmetries of the molecule.

\begin{figure}
\includegraphics[width=0.75\columnwidth]{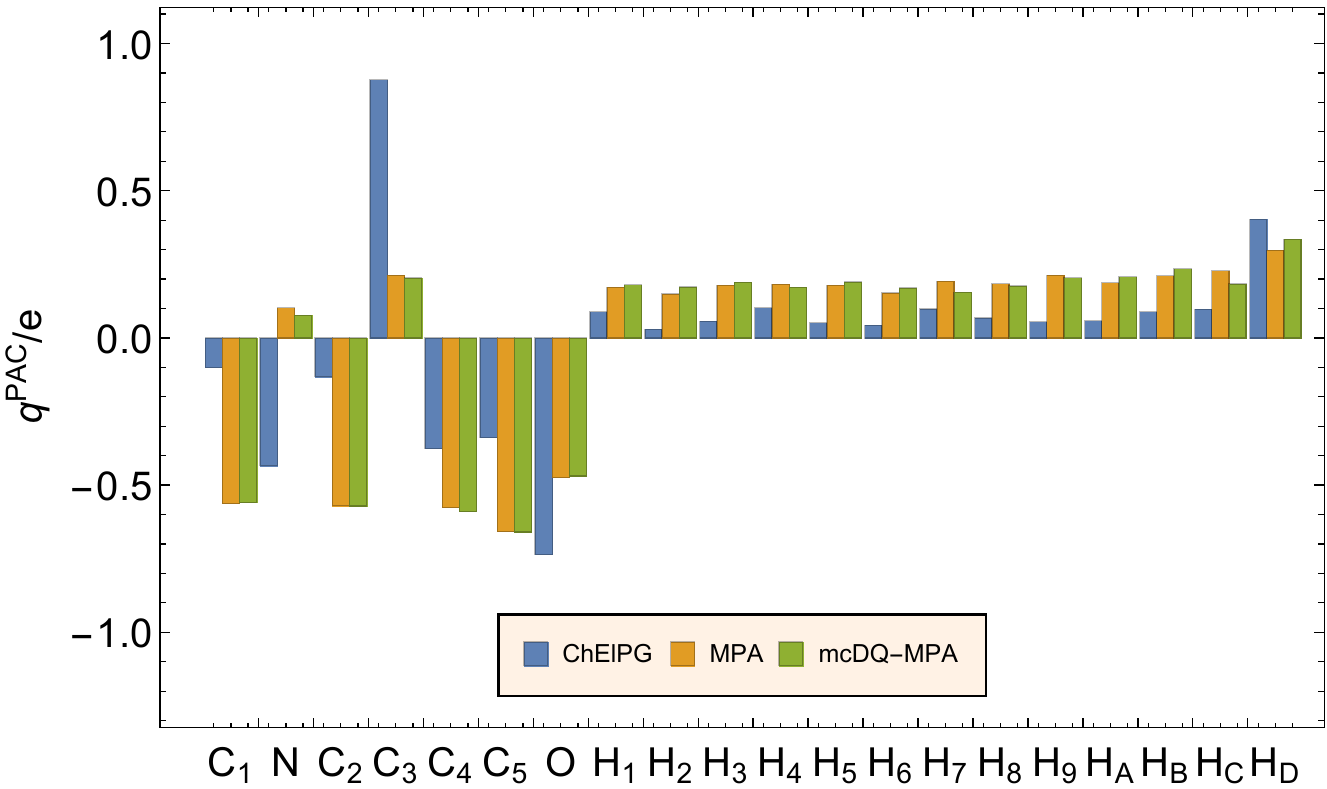}

\caption{\label{fig:Burial}The ChElPG, MPA and mc-MPA PACs for the 2-(Dimethylamino)-2-propanol }
\end{figure}

An additional benefit of the mc-PACs is their stability for inner
(or buried) atoms of large molecules. This rises from the stability
of the standard population schemes themselves and the fact that mc-PACs
involve rather small corrections. As an example, consider the 2-(Dimethylamino)-2-propanol
molecule:

\noindent\begin{minipage}[t]{1\columnwidth}%
\begin{center}
\includegraphics[width=0.4\textwidth]{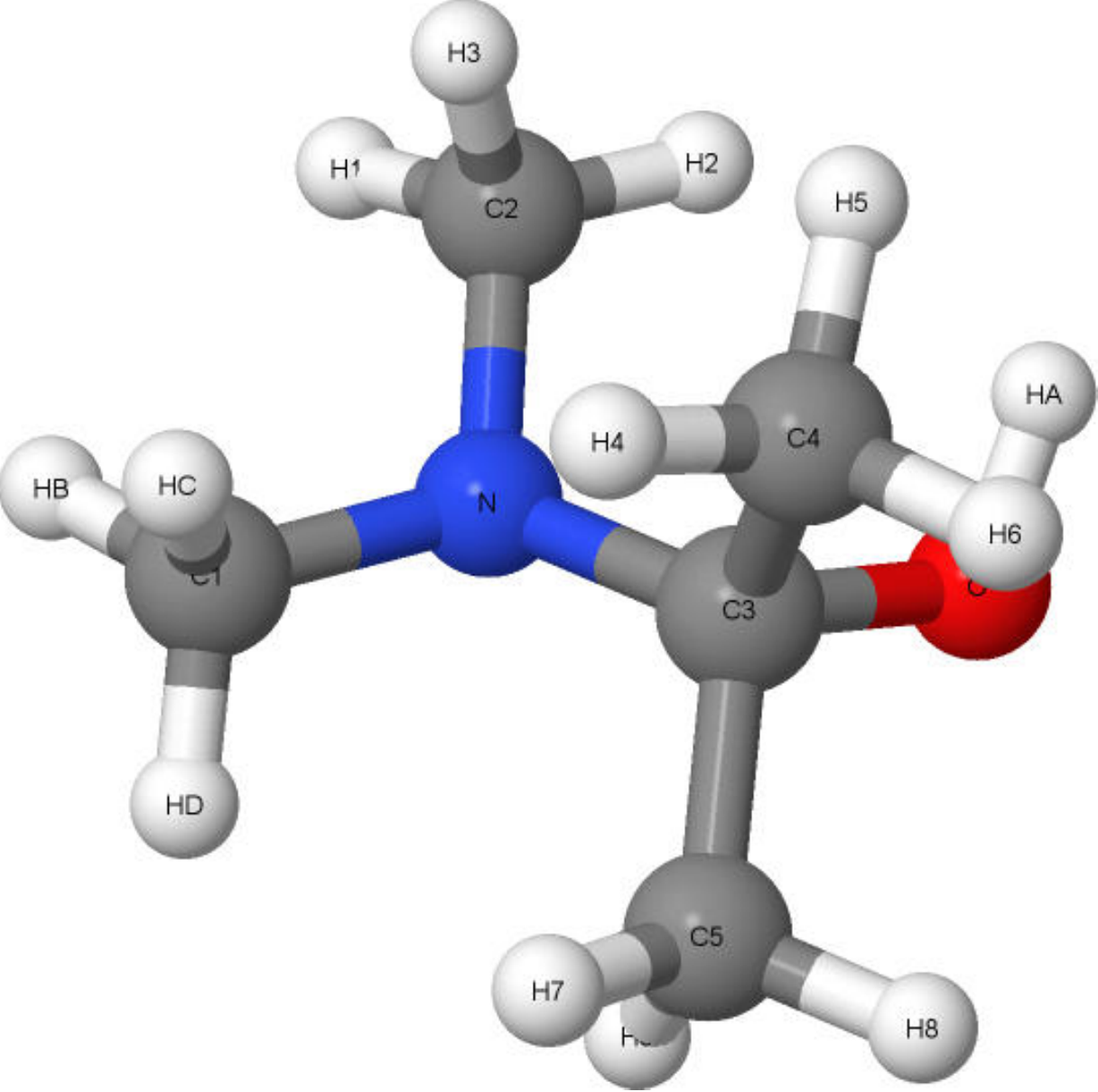}
\par\end{center}%
\end{minipage}\\
for which the ChElPG, HPA and mcDQ-HPA PACs are shown in Fig.~\ref{fig:Burial}.
Here, ChElPG tends to polarize the molecule: the oxygen and nitrogen
share between them a negative unit charge and this is counteracted
by the positive unit charge of the central carbon atom C3. On the
other hand, MPA assigns a low charge for C3 and spreads rather evenly
the remaining positive charge on the 12 terminal hydrogen atoms. mc-MPA
charges are very close to those of MPA and thus yet they improve significantly
the ESP description for this molecule: the MPA MARD is 0.35 while
that of the mc-MPA is 0.1. It is worthwhile to note that the PACs
assigned by ChElPG also have a MARD of 0.1.

When the underlying reference is the iHPA set of PACs the resulting
ESP is of similar quality to that of the ChElPG set of PACs resulting
from a best-fit to ESPs. The dependence of the PACs on the molecular
distortion was demonstrated to have sometimes very sharp features
however all the changes were smooth, hence forces can be calculated
on the atoms of the molecule.

The method here bears a similarity to the optimal point-charge model
of Ref.~\cite{Simmonett2005} which determines PACs that reproduce
as many low-order moments as possible. The crucial difference is best
seen when systems grow, model of Ref.~\cite{Simmonett2005} would
target increasingly higher electrostatic moments as more atoms are
included while the present method targets multipoles up to second
order and not beyond, thereby avoiding the numerical instabilities
described in see Ref.~\cite{Gilbert2006}. On the other hand. the
optimal point-charge model treats the multipole constraints in a more
systematic way by minimizing the error over unused moments in the
last incomplete spherical shell.

\section*{Acknowledgments}

Authors express special thanks to Dr. Yihan Shao from Q-CHEM Inc.
for his advice and critical assistance in performing the iHPA calculations.
We also gratefully acknowledge the support of the Israel Science Foundation
Grant No. 189/14. 

\bibliography{EfratBibLib}

\end{document}